\documentclass{aa}
\usepackage{natbib}
\usepackage{graphicx}
\usepackage{txfonts}

\begin{document}
\title{Dispersal of planets hosted in binaries,
transitional members of multiple star systems}
\subtitle{}

\titlerunning{Planets in binary systems}

\author{F. Marzari
        \inst{1}
        \and
        M. Barbieri
        \inst{2}
        }

   \offprints{F. Marzari}

   \institute{
              Dipartimento di Fisica, University of Padova, Via Marzolo 8,
              35131 Padova, Italy\\
              \email{marzari@pd.infn.it}
         \and
             LAM, Traverse du Siphon, BP 8, Les Trois Luc,
             13376 Marseille Cedex 12, France\\
             \email{mauro.barbieri@oamp.fr}
             }

   \date{Received XXX ; accepted XXX}

\abstract 
{}
{If a significant fraction of binary star systems spent some time as an
inclined triple, either during their formation process or as the outcome
of several close dynamical encounters in a crowded stellar environemnt, then 
the number of planets in binaries would be signifcantly lower 
than that around single stars. 
The stellar chaotic phase preceding the instability of the 
triple
and the wide oscillations in eccentricity and inclination of 
the companion star due the high mutual inclination between the 
companion and the singleton  would quickly eject planets orbiting the 
binary in S--type orbits.  
}
{We perform numerical simulations of the dynamical evolution of 
hierarchical triple star systems with planets hosted around the primary
star of the inner binary. Different values of mutual inclination, 
binary separation and singleton initial semimajor axis are
explored in a statistical way.}
{We find that a significant mutual inclination
$i_m$ between the singleton and 
the binary is a key factor for instability of the planetary 
system. When $i_m$
is larger than $\sim 40^{\circ}$ the fraction of planets 
in the binary
surviving the chaotic phase of the triple declines dramatically. 
The combination of 
eccentricity and inclination oscillations of the binary companion
induced by the secular perturbations of the singleton 
and the sequence of close encounters preceding the ejection 
of one star fully destabilize a planetary system 
extending beyond 1 AU from the star. 
For $i_m$ around $90^{\circ}$  the percentage of surviving planets
is lower than 20\% for all binaries with a semimajor 
axis smaller than 200 AU.}
{The frequency of planets in binaries with low separation may 
be strongly reduced by the residence of the pair in the past  
in a temporary inclined hierarchical triple.  
}


\maketitle

\section{Introduction}

Over 65 percent of the main sequence stars in the solar neighborhood 
are members of binary or multiple star systems (Duquennoy and Mayor, 1991).
As a consequence, 
answering the question of whether planets can form and persist near one
of the stars in a
binary has far-reaching implications for the overall frequency of 
planetary systems. Studies on the long term stability of planets in binaries 
have shown that a planet cannot be located too far away from
the host star or its orbit will be destabilized by the 
gravitational perturbations of the 
companion star.   
Holman and Wiegert (1999) 
found that the stable/unstable boundary depends on the mass ratio and 
eccentricity of the binary, but for a wide range of parameters 
stable orbits may extend well beyond one tenth of the binary
semimajor axis. However, in terms of the probability of finding a planet 
in binary systems the dynamical stability analysis is not exaustive 
since it does not take into account the profund influence that 
stellar dynamic interactions may have had on the early evolution of
a planetary system in a binary. 

It has been suggested that most binaries
originate from the decay of multiple
systems (Reipurth, 2000; Larson 1972, 1995, 2001 ; Kroupa 1995).
The most common configuration among multiple systems is the hierarchical 
triple, where a singleton orbits around the baricenter of a binary system.
A hierarchical triple can become unstable after some time,
depending on its initial orbital parameters, leading to the 
disintegration of the triple \citep{egki,egki2}.
This disintegration occurs via a
phase of chaotic evolution whose outcome is the ejection of
one of the three stars (typically the least massive body) on
an unbound trajectory. The other two stars, members of the 
original binary, are left in a more tightly bound binary.
In a previous paper (Marzari and Barbieri, 2007, herein after 
MB1) we showed that
the orbital changes of the binary and the strong gravitational 
perturbations during the chaotic phase prior to the singleton 
ejection can influence the final configuration of a planetary 
system hosted by the primary star of the pair. 
However, in the context of near--coplanarity between the binary and the 
singleton, planets can survive the triple decay in most cases  
and adapt to the new orbital parameters of the binary. 
The major effect would be a significant change in the orbital 
configuration of the system after the triple instability 
with respect to the original configuration, as an outcome of the 
planet formation process. 

In this paper we consider the dynamic effects of the 
decay of {\it inclined} 
hierarchical triples on planetary systems. 
In particular, we will 
focus on planet survival during the unstable triple 
configuration. 
At present, determinations of the mutual inclinations 
of the two orbits 
in hierarchical triple stellar systems are available  
only for a very limited number of cases and are often
ambiguous.
\cite{feke} determined that at least 1/3 of a sample of
20 
triple star systems have an inclination exceeding
$15^{\circ}$ and are not coplanar.  
\cite{stto} analysed a different set of 22 visual triples
finding an average mutual inclination of
$79^{\circ}$.  However, in both the studies the 
mutual inclination was derived from incomplete 
observational data.  To compute unambiguous mutual inclinations 
for triples, both radial velocities and visual orbits are 
required
for the inner and outer system. 
So far, only six nearby systems have been observed with both methods and 
have direct and precise measured orbits
\citep{muto}. The values of mutual inclination for
these systems range from
$24^{\circ}$ to $132^{\circ}$ but the sample is too small
to give 
hints on the real 
distribution of inclinations among triples.  
The mutual inclination of triples may either be  
primordial and related to the
formation process of the triple by fragmentation of a molecular cloud
or it may form 
at later times because of 
dynamical interactions, like encounters, between single stars and binaries
in a dense cluster-like environment. In the latter case the inclination
is due to the encounter geometry between the binary and the single 
star and should be randomly distributed. 
Any deviation from randomly oriented orbits may be an important indication
of the relative importance of the two formation
mechanisms.
Assuming that planets can form
in the binary before the bound hierarchical triple becomes unstable, 
the dynamical interactions between the stars during the chaotic 
phase can strongly 
affect the stability of the planetary system. 

We can envision two different scenarios for planet formation
and subsequent destabilization within an inclined hierarchical 
triple: 

\begin{itemize}

\item A primordial binary star system forms in a star cluster 
and planets accumulate from a circumstellar disk around 
the main star either by core--accretion \citep{pollo} 
or by disk instability \citep{bos1}. 
The existence of a few 
gas giant planets in binary systems with separation
of a few tens of AU ($\gamma$ Cephei and GL 86) 
suggests that the perturbations of the 
companion star are not 
strong enough to prevent the formation of
planets in binaries \citep{the06, boss}. Successively, a
temporary hierarchical triple builds up because of dynamical interactions 
between the primordial binary, with planets, and a passing by singleton or binary 
\citep{ford}. In a dense stellar environment with a large
abundance of binaries this is a 
frequent event \citep{mcmi}. 
In the presence of a significant inclination between the singleton 
and the binary orbit,
the planetary system in the binary is 
strongly destabilized during the transitional triple state 
by both the 
secular perturbations of the singleton 
and the frequent stellar encounters 
during the chaotic phase preceding the break up of the 
triple into a binary and singleton. 
The destruction of the triple may occur either because it is 
unstable and after a short timescale the singleton 
escapes or because of an encounter with other objects, single
stars or binaries.
After the ejection of the
singleton in a hyperbolic orbit, the primordial binary has
different orbital parameters but it is also 
depleted, in most cases, of its original planetary system. 

\item  A primordial inclined hierarchical triple forms by fragmentation of 
a single, rotating, dense molecular cloud \citep{bosl}. 
Planets can grow on S--type orbits around the 
primary star of the binary, possibly by disk instability \citep{bos1}.
Self-gravitating  density
clumps can contract into planets in only a few hundred years.
It is 
unrealistic to expect that planets can form by core--accretion
in this scenario 
because of the strong secular perturbations
that the singleton 
would apply on a long timescale on the secondary star and then, 
indirectly, to  
a putative planetesimal disk around the binary
(assuming that planetesimals
could form in such a highly perturbed circumstellar disk). 
Even in this case, the combined destabilizing effects of
secular perturbations and stellar encounters in the
chaotic phase preceding the ejection of the
singleton in a hyperbolic orbit destabilize the planetary
system around the primary. 
\end{itemize}

In this paper we numerically model the orbital evolution of 
planets in S--type orbits in a binary member of an inclined 
unstable hierarchical triple.
We consider different mutual inclinations
between the binary and the singleton, while in MB1 we simulated 
only the planar case. We find that, 
contrary to the low inclined cases, only in a 
limited number of cases do
planetary systems extending beyond 1 AU survive 
after the chaotic phase 
of stellar encounters when the mutual 
inclination between singleton and companion is larger than 
$\sim 40^{\circ}$. There are two mechanisms that,
acting in synergy, destabilize planets
around the primary star.
\begin{itemize}

\item A large initial mutual inclination 
between the outer stars excites consistent eccentricity--inclination oscillations 
of the binary companion with periods of the order of some thousand years 
and more. These oscillations, well described by a quadrupole--level 
secular theory \citep{maz,ford},  strongly
affect the orbits of the planets around the primary, forcing most of 
them to leave
the system on hyperbolic orbits. In the phase of high eccentricity, 
the companion moves closer to the planetary system, reducing
the region of stability \citep{howi} and perturbing the planetary orbits.
After a few Kozai--cycles a large fraction of the planetary 
system is destroyed. 
For retrograde orbits the dynamics is 
more complex but the evolution is still characterized by wide 
oscillations of both eccentricity and inclination. 
This kind of Kozai mechanism in stellar triples is  
different to that described by \cite{melma} where  
the secular interactions involved only the 
companion star of an isolated binary and the planets. 

\item The second mechanism giving the 'coup de grace' to the planetary system 
is the sequence of close encounters between the singleton and the 
companion star of the binary occurring during the chaotic phase 
preceding the disgregation of the triple. Close encounters 
between stars on mutually inclined orbits are much more effective 
in destabilizing the planets than in the low inclination case. 
\end{itemize}

Both the eccentricity
oscillations  of the 
secondary star and the close encounters conspire against the 
survival of a planetary system around the main star of the binary. 
If indeed binary stars
are born as triple or higher multiplicity stellar systems or they 
are temporarily involved in unstable triples, their planetary
systems would be fully destabilized in most of the inclined cases.  
The fraction
of planets in binaries observed at present 
would than be lower than that 
around single stars. The number
 of binaries depleted of 
planets would depend on 
the distribution of  
mutual inclination in the primordial triples. 
If coplanarity dominates, as might be the case if 
the majority of temporary triples originated directly
from the fragmentation of an interstellar cloud \citep{bosl}, then 
the binaries stripped of their planets would be a 
minority. On the other hand, if most of the triples 
formed by gravitational interactions in a dense 
stellar environment, the large mutual inclinations
would lead to a strong planet depletion among the 
surviving binaries. In this scenario, it is an important 
observational challenge to increase the statistics of 
known triple systems with unambigous determination of the 
mutual inclination.  
At present only about 15\% of planets have 
been found in multiple stellar systems.  This is probably 
an observational selection effect in favor of single
target stars but it might also be a first  
indication that binary systems are depleted 
of planets by the past violent dynamical evolution of the stars.

We will not explore in this paper the full complexity of the 
hierarchical triple dynamics as performd in 
\cite{ford}. We are interested on the consequences 
of the large variations of the star orbital elements 
on the planets and we perform statistical 
numerical simulations giving the fraction of planets surviving 
the chaotic phase of unstable triples. We also do not 
investigate the planetary formation process in detail, but
we assume that planets can form by either of the two mechanisms,
core--accretion or gravitational instability. 

In Sec. 2. we describe the numerical model adopted for the 
numerical integration of the trajectories of the stars and planets. 
Sec. 3 is devoted to the statistical analysis of the 
survival of planets in S--type orbits around the primary star. 
In Sec. 4 we present our conclusions. 

\section{The model}

Our numerical model consists of 3 stars, two locked in a binary system
and the third orbiting the barycenter of the pair. A  set of 10 massless
bodies started on circular orbits around the primary star simulate a 
planetary system that formed in the early phases of evolution of 
the binary. The semimajor axes of the test bodies are regularly spaced 
from 1 to 10 AU and the initial inclinations are all set to $0^{\circ}$
with respect to the binary orbital plane. The trajectories of the stars and of 
the 'planets' are computed with the numerical integrator RADAU \citep{rad}.
It handles in a very precise manner close encounters between massive bodies
and it does not require a fixed hierarchical structure such as HJS 
\citep{beu} or SYMBA5 \citep{dunc_levi}.  

To model the outcome of the triple instability in all possible configurations
is a difficult task since the parameter space to explore is 
wide. For this reason we select a limited number of parameters
to be sampled while the others are left unchanged. To better compare
our results with those presented in MB1, we adopt the same masses for 
the stars i.e.  1 and 0.4 solar masses for the binary, 0.4 solar mass
for the singleton. An eccentricity of 0.2 is adopted for both the binary 
and the singleton, taking into account that the orbit of the 
singleton is defined with respect to the barycenter of the binary.  
The mutual inclination is sampled between $0^{\circ}$ and 
$180^{\circ}$ including in this way retrograde orbits of the 
singleton. For any value of the semimajor axis of the binary
$a_b$, we sample
different values of the semimajor axis of the singleton $a_s$ and 
of the orbital angles other than those giving the mutual 
inclination. For any set of $(a_b, a_s, i_m)$ 
we perform 20 simulations with random initial orbital 
angles to increase the statistics on the star and planet 
dynamical behaviour.  

\section{The dynamical sources of instability}

\begin{figure}
\resizebox{15cm}{!}{\includegraphics{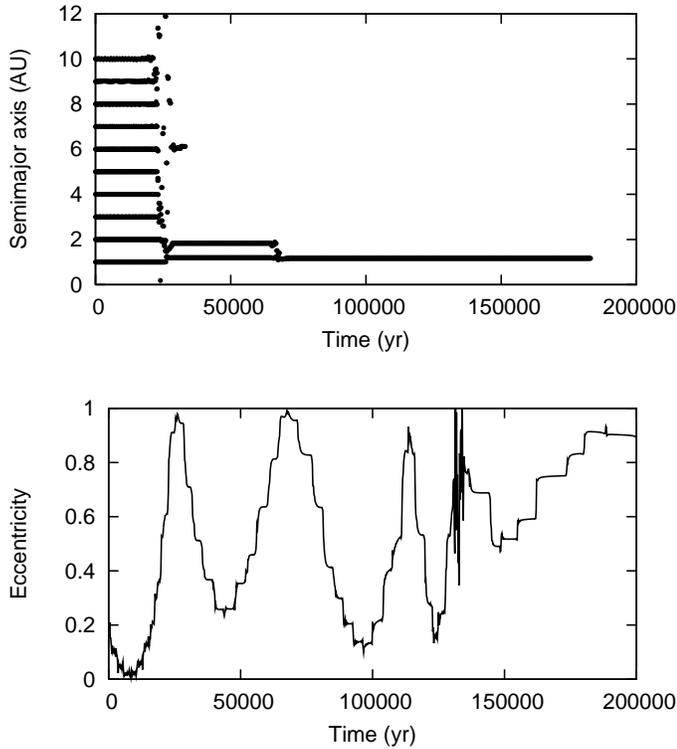}}
\caption[]{Orbital evolution of the planets around the 
primary star of the binary (plot a) under the perturbations of the
companion star, in turn affected by the gravitational 
pull of the outer singleton star (plot b). The initial 
semimajor axis of the binary is 70 AU, the eccentricity
of the binary 0.2, that of the singleton 0.2, and the mutual 
inclination $i_m$ is set to $90^{\circ}$.
}
\label{f1}
\end{figure}

In this section we discuss in detail the two mechanisms leading to destabilization 
of a putative planetary system extending beyond 1 AU around the primary star 
of a binary in an inclined temporary triple. 
In Fig.\ref{f1} we show the evolution of a model with $a_b = 70$ AU, 
$a_s = 212 $ AU and initial mutual inclination $i_m = 90^{\circ}$. 
In this configuration, the critical semimajor axis for 
long--term stability of planetary orbits around the primary is, 
according to \cite{howi}, around 21 AU. Our initial planetary system, 
extending out to 10 AU, is well within the stability region. 
The perturbation of the singleton induces
Kozai cycles on the binary companion that achieves an
eccentricity of almost one  
over a timescale of $2.5 \times 10^4$ yrs. 
This behaviour is well described by quadrupole and octupole--level
secular equations described in \cite{maz,ford}.
All the planets beyond 
2 AU are ejected from the system after the first cycle, while that orbiting
at 2 AU is destabilized after the second cycle. Starting from
$1 \times 10^5$ yrs the singleton and the companion star have
mutual close encounters that quickly lead to the ejection of the 
last inner planet, that lived through the Kozai cycles of the companion. 
Finally, after about 
$5 \times 10^5$ yrs, the outer star is ejected on a 
hyperbolic orbit and the the binary system is left with a
smaller separation but no planets. 

This kind of behaviour, typical of systems with high mutual inclination
$i_m$, 
places in jeopardy not only the stability of planets around the primary 
but also the possibility that they can form.  According to \cite{bos1}, 
several gaseous protoplanets can rapidly form by disk instability
in a marginally gravitationally unstable protoplanetary disk. Within this 
scenario 
in a few hundreds years we might witness the formation of the 
unstable triple and of a planetary system made of gas giant planets  
around the primary before the Kozai cycle increases the eccentricity
for the companion star. On the other hand, core--accretion would 
not have enough time to accumulate a core by planetesimal accretion, and even 
planetesimals may have failed to form on such a short timescale. 
A protoplanetary disk around the primary star would be strongly perturbed 
and almost fully 
destroyed during the first Kozai cycle in eccentricity and 
inclination of the companion star. However, 
if the binary system was isolated during its formation and it became part of  
an unstable triple later on because of repeated stellar encounters
in a dense star--forming region, then 
planets might have the time to grow even by core--accretion, before 
the onset of the strong perturbations related to the stellar interactions
in the triple phase. 

\begin{figure}
\resizebox{15cm}{!}{\includegraphics{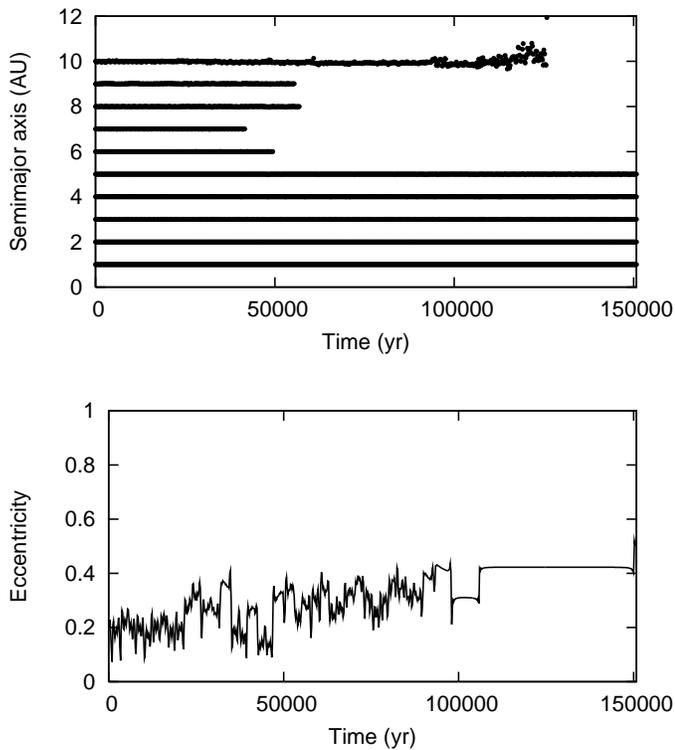}}
\caption[]{Same as in  Fig.\ref{f1} but for a lower value of 
$i_m$ set to $30^{\circ}$.
}
\label{f2}
\end{figure}

Only systems with large values of $i_m$ are fully destabilized by 
the stellar perturbations of the triple. 
When the mutual inclination is lower than $\sim 40^{\circ}$,
close encounters between the stars, and the consequent impulsive changes
of the orbital elements, are a source of instability for the 
planets but often not strong enough to destabilize the full 
planetary system. 
In Fig.\ref{f2} we illustrate the evolution of a model with $a_b = 70$ AU,
$a_s = 212 $ AU, as in the previous case, but with a lower initial mutual 
inclination $i_m = 30^{\circ}$. The triple quickly becomes unstable and 
the singleton has frequent close approaches with the binary companion 
marked by sudden steps in eccentricity and semimajor axis. The 
changes in the orbital elements of the companion leads to unstable planetary
orbits as shown in Fig.\ref{f2}. However,  contrary to the case shown in 
Fig.\ref{f1} the planetary system is not fully destroyed and planets 
within 5 AU of the star survive the chaotic phase.
Further perturbations by the binary companion  
after the triple disruption do  
not destabilize the planetary survivors since they are well
within the critical semimajor axis for stable orbits \citep{howi}.

\begin{figure}
\begin{center}
\begin{tabular}{c}
\resizebox{90mm}{!}{\includegraphics{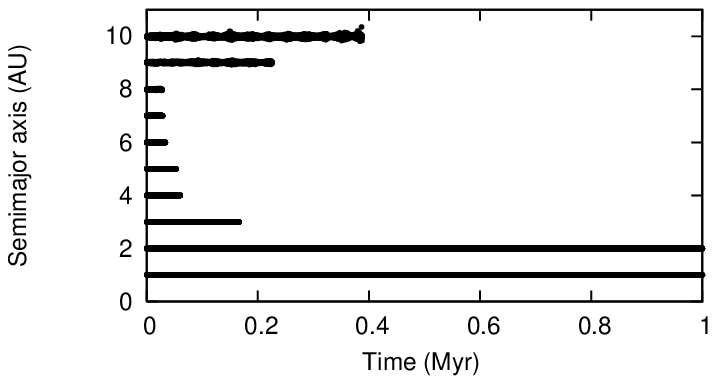}} \\
\resizebox{90mm}{!}{\includegraphics{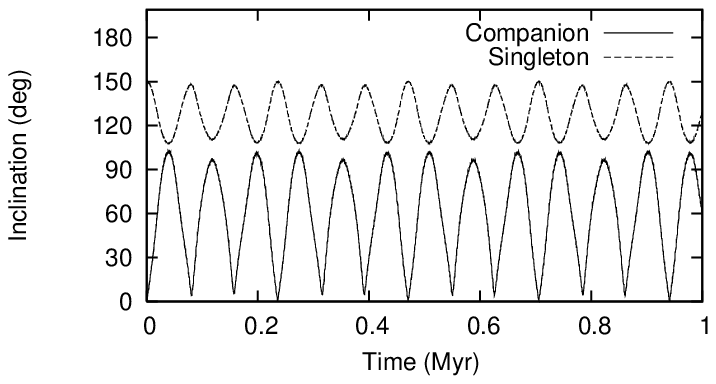}} \\
\resizebox{90mm}{!}{\includegraphics{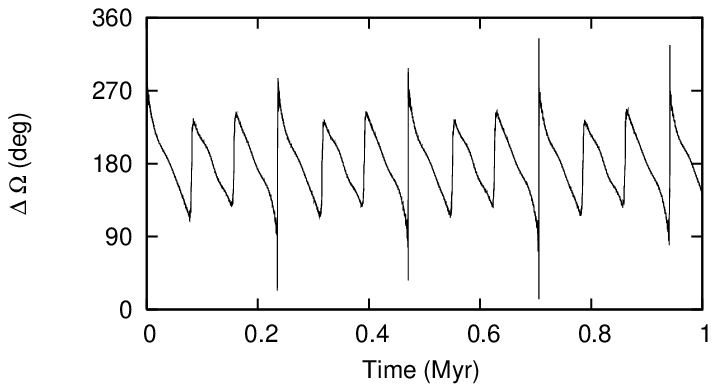}} \\
\end{tabular}
\caption[]{Destabilization of planets around the primary 
when the singleton is on a retrograde orbit relative to the 
companion.  The initial mutual inclination between the 
two outer stars is $150^{\circ}$. The $\Delta \Omega$ angle 
librates around $180^{\circ}$ while the inclinations of the
two planets have wide oscillations. Only two planets, the 
closer ones, survive after 1 Myr of evolution. 
}
\label{f3}
\end{center}
\end{figure}

If the companion and the singleton are on retrograde orbits, instability
builds up in a similar way. For mutual inclinations lower than 
$\sim 140^{\circ}$, large amplitude oscillations of the eccentricity 
begin to destabilize the planetary system which is finally 
destroyed by the stellar encounters in the chaotic phase. Contrary to 
the prograde case, the oscillations of eccentricity and  
inclination are not in phase, as predicted by the quadrupole theory. 
Apparently, the two orbital parameters are no longer bound in an 
invariant and they evolve with independent frequencies. For inclinations
in between $\sim 140^{\circ}$ and $\sim 180^{\circ}$ the oscillations 
in eccentricity are moderate to low but some instability of planetary orbits 
is driven by the large inclination oscillations of the companion. 
In Fig.\ref{f3} we show the evolution of the inclinations of the 
singleton and companion star when the mutual initial inclination is 
$\sim 150^{\circ}$. We plot the inclination of each individual
star, referred to the initial plane of the binary, because this is
the plane where the planets also begin to orbit the primary. 
The inclination of the companion becomes very high and 
becomes retrograde for a short while. The behaviour is characterized also
by the libration of the angle $\Delta \Omega_2 - \Delta \Omega_1$ 
(see  Fig.\ref{f3}, lower panel) with the same frequency as the inclination 
oscillations. Most of the planetary orbits are destabilized during these
large inclination excursions of the companion star and after 1 Myr only
the two inner planets survive. The onset of the chaotic phase of 
the stars ejects finally also these two survivors. 

\section{Statistical outcome}

\begin{figure}
\resizebox{\hsize}{!}{\includegraphics{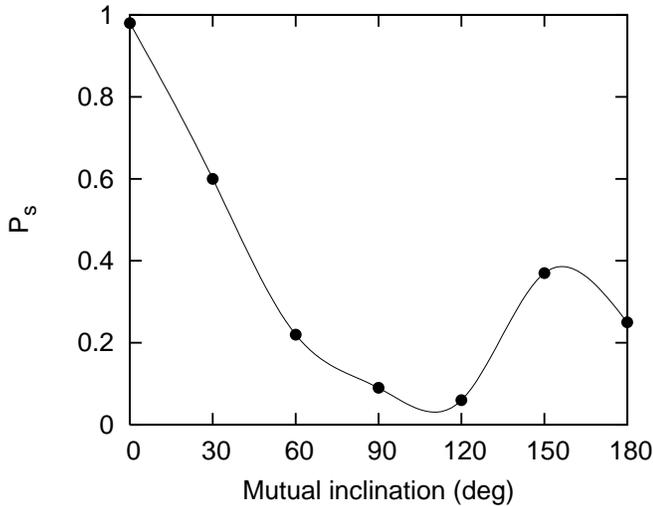}}
\caption[]{Fraction $P_s$ of binaries that, at the end of 
the temporary triple phase, retain at least one 
of the planets in a stable orbit around the 
primary vs. $i_m$, the initial mutual inclination 
between the singleton and the binary companion.
}
\label{f4}
\end{figure}

To test the chances of a planetary system in a binary 
to survive a period of stellar interactions typical of 
an unstable triple, we have run several models with
the binary semimajor axis fixed to 
$a_b = 70$ AU. The orbital eccentricities of the stars are both set
to 0.2. In Fig.\ref{f4} we plot the percentage $P_s$ of 
dynamical systems that, at the end of the period as a
hierarchical triple, retain at least one of the initial 10 planets
vs. $i_m$, the initial mutual inclination between
the two outer stars. This percentage is very high for low inclinations
confirming the results presented in MB1 for low--inclination 
systems, 
while it declines very quickly when the inclination approaches $90^{\circ}$.
This is a consequence of both the Kozai cycle that pushes the 
binary companion closer to the planets, and of the more complex 
orbital behaviour during close approaches between the stars when 
their orbits are inclined. 

Retrograde orbits of the singleton also lead to 
fast instability of the planets when the mutual inclination is 
close to $90^{\circ}$. Wide oscillations of the eccentricity up 
to large values are observed, even if not related to the known 
Kozai type mechanism: there is no phasing between eccentricity 
and inclination.   However, even in this case when
the eccentricity is at its peak value most of the 
planets are destabilized.  Only when the mutual inclination has values 
beyond $\sim 140^{\circ}$ the planets around the primary 
are partly spared by the oscillations in eccentricity of the binary 
companion. However, as observed in the previous section, 
for mutual inclinations in the range $\sim 140^{\circ} - 180^{\circ}$ 
large inclination oscillations of the companion star destabilize 
planets even if to a lesser extent than the
eccentricity oscillation. As a result, 
the percentage of planets surviving the chaotic phase grows 
for inclinations larger than $90^{\circ}$ but it does not
return to 100\%, halting at about  
30\%.

\begin{figure}
\resizebox{\hsize}{!}{\includegraphics{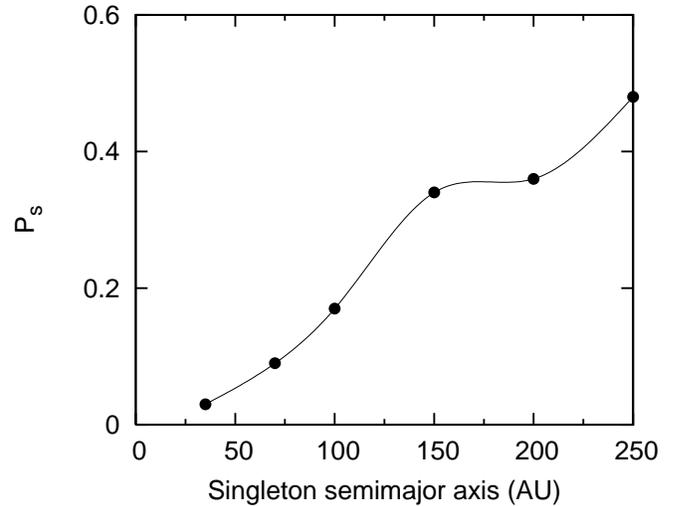}}
\caption[]{Fraction $P_s$ of binaries that, at the end of 
the temporary triple phase, retain at least one 
of the planets in a stable orbit around the 
primary vs. $a_b$, the binary semimajor axis. 
The mutual inclination $i_m$ is set to $90^{\circ}$,
the worst case for planet stability.
}
\label{f5}
\end{figure}

If we increase the semimajor axis of the binary $a_b$, 
the fraction of systems with surviving planets
increases in an almost 
linear way. In Fig.\ref{f5} 
we show the fraction of systems retaining planets vs. $a_b$
for the worst case, i.e. with mutual inclination equal to
$90^{\circ}$. 
The triple instability is a mechanism that easily destroys
planetary systems of close binaries while it is less effective for 
wide binaries. For larger values of $a_b$ the planetary systems 
that survive are also more extended in semimajor axis. 
In most cases for $a_b = 250$ AU all the planets up to 
$a_p = 10$ AU survive the stellar chaotic phase.

\section{Conclusions}

The fraction of binary systems hosting planets in S--type orbits can be 
lower than expected. If the binary is part of 
a crowded
stellar environment, encounters with other stellar objects 
can lead to the formation of a transitional 
triple with large mutual inclination between the singleton and 
the binary. The subsequent dynamic evolution of the triple, 
in particular the large oscillations in eccentricity of the 
companion star in the binary and the chaotic evolution 
during the triple destruction, destabilize planetary orbits around 
the main star. Even if the binary was born as part of an unstable 
inclined triple, the planetary system is fated to be disrupted. 

Observing a binary system without planets in S--type orbits does
not necessarily imply that the stars did not posses circumstellar
disks in their early phases or that planet formation did not 
occur. The history of the binary and of its primordial 
environment must be taken into account since in most cases 
it may be the cause of the absence of planets. Planet formation
might be a very efficient process also in the presence of 
external perturbations, but the survival of planetary systems 
may be threatened by the binary dynamical history. 

\begin{acknowledgements} We thank P. Eggleton for stimulating us 
to perform this work.
\end{acknowledgements}

\end{document}